%% file: main.tex
\documentclass[11pt]{article}
\usepackage[preprint]{acl}
\usepackage{times}
\usepackage{latexsym}
\usepackage[T1]{fontenc}
\usepackage[utf8]{inputenc}
\usepackage{microtype}
\usepackage{inconsolata}
\usepackage{booktabs}
\usepackage{graphicx}
\usepackage{placeins}
\usepackage{amsmath,amssymb}

\title{SEAL: Sealed Queries for Local Sanitization\\
against Indirect Prompt Injection}
\author{Doohyun Kim\textsuperscript{1} \quad Donghwa Kang\textsuperscript{2} \quad
Mingyu Jun\textsuperscript{2} \quad Brent ByungHoon Kang\textsuperscript{2} \\[4pt]
\normalsize\textsuperscript{1}Graduate School of Information Security, KAIST \quad
\textsuperscript{2}School of Computing, KAIST \\
\normalsize Daejeon, Republic of Korea \\[2pt]
\normalsize\texttt{\{doohkim7, anima0729, mingyu.jun, brentkang\}@kaist.ac.kr}}
\begin{document}
\maketitle
\input{sections/abstract}
\input{sections/introduction}
\input{sections/related}
\label{draft:sections1-2-end}
\input{sections/observations}
\label{draft:sections1-3-end}
\input{sections/method}
\label{draft:sections1-4-end}
\input{sections/evaluation}
\label{draft:sections1-5-end}
\input{sections/conclusion}
\label{naacl:body-end}
\clearpage
\input{sections/limitations}
\input{sections/ethical_considerations}
\label{draft:references-start}
\bibliography{references}
\clearpage
\appendix
\input{appendix/experimental_setting_details}

\input{appendix/observation_details}
\input{appendix/evaluation_details}

\end{document}

%% file: sections/abstract.tex
\begin{abstract}
Indirect prompt injection threatens local LLM agents through untrusted
tool outputs, motivating defenses that suppress attacks while preserving
task evidence under limited inference budgets. We introduce SEAL, which
learns a soft prompt and scoring readout offline to sanitize tool outputs
using a frozen task model. SEAL constructs scoring queries from trusted
initial context and the learned prompt, keeping the reference independent
of inspected tool text. It scores tokens without normalization across
context tokens and uses preservation regularization to encourage retention
of useful evidence during localized deletion. This design requires neither
defense-generated tokens nor a separate verifier model. Across
AgentDojo and AgentDyn, SEAL suppresses attacks while improving task
completion under attack. Specifically, on AgentDojo, SEAL reduces attack
success from 34.6\% to 0.1\% and increases Joint---task completion without
attack success---from 51.4\% to 77.5\%, while using approximately
$127\times$ fewer defense-related tokens than PISanitizer, a state-of-the-art
prompt injection defense. Replication on a second backbone with the same
training recipe supports applicability across model families.
\end{abstract}

%% file: sections/introduction.tex
\section{Introduction}
\label{sec:intro}
Recent advances in compact, openly available large language models
(LLMs) are making local deployment increasingly practical, improving
reasoning and instruction following at modest model sizes
\citep{gemmateam2025gemma3,qwen2026qwen35}.
Developers report concerns about agent data security and privacy
\citep{stackoverflow2025survey}, and local LLM pipelines already support
privacy-preserving clinical information extraction \citep{wiest2024privacy}.
Improving capabilities and practical demand for data control make
securing local agents increasingly important.

\input{figures/defense_families}

Local deployment does not make the content an agent reads trustworthy.
A retrieved page, message, or record may contain useful task data alongside
instructions written to redirect the agent. This threat, \emph{indirect
prompt injection} (IPI), crosses the boundary between external data and
authorized instructions
\citep{greshake2023not,zhan2024injecagent,debenedetti2024agentdojo}.
We consider local agents whose initial instructions and tool definitions
are trusted but whose tool outputs may be attacker-controlled. We keep
the task model frozen, protecting the model the operator has already selected.

Memory capacity and model-loading costs constrain local inference
\citep{alizadeh2024llmflash}, while multiple large models can incur memory
contention and transfer overhead on consumer GPUs \citep{xu2026nixie}.
We therefore target resource-constrained deployments, seeking stronger
security with preserved task performance and limited additional inference,
without requiring a separate verifier LLM.
Figure~\ref{fig:defense-families} summarizes three defense families that
reuse the selected task model.

These families present different tradeoffs for our goal.
\emph{Natural-language reinforcement} is lightweight
\citep{learnprompting2024sandwich,hines2024spotlighting}, but leaves
substantial attack success in our local-agent evaluation
(\S\ref{sec:eval-main}). \emph{Auxiliary LLM inference} can strengthen
protection \citep{he2026attriguard,li2025drift}, but extra calls and
generated decisions consume additional inference budget even when they
reuse the task model (\S\ref{sec:eval-cost}).

\emph{Attention-based defenses} offer a promising route by using internal
attention signals, reducing reliance on generated verification responses
\citep{hung2025attentiontracker,geng2026pisanitizer}.
The challenge is to make these signals reliable without sacrificing task
evidence: scores can depend on the inspected text and surrounding context,
and removal can discard useful content. These challenges motivate the
design requirements in \S\ref{sec:observations}.

We introduce \textbf{SEAL}, which searches for a short \emph{soft prompt}
---a sequence of continuous input embeddings---that helps the frozen
model distinguish injected instructions from benign task content.
SEAL combines a scoring reference built from trusted context with
selective removal optimized to preserve useful content. It sanitizes
tool outputs before the agent continues, without generating defense
responses or changing the task model's weights.

Our contributions are:
\begin{itemize}
    \item We analyze attention-based sanitization in frozen local agents
    and identify requirements for reliable scoring and task preservation.
    \item We introduce SEAL, which searches for a soft prompt to elicit
    security signals from the task model and selectively removes injected
    instructions while preserving benign content.
    \item Through evaluations against seven baseline defenses across
    multiple benchmarks and additional robustness analyses, we demonstrate
    that SEAL reduces attack success while maintaining high Joint with
    low defense-related token costs.
\end{itemize}

%% file: figures/defense_families.tex
\begin{figure}[t]
  \centering
  \includegraphics[width=\columnwidth]{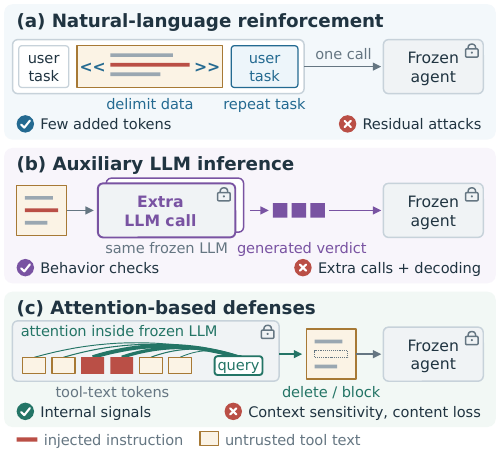}
  \caption{Representative defense pathways for frozen local agents.
  (a) Natural-language reinforcement adds reminders or delimiters to the agent input.
  (b) Auxiliary LLM inference generates extra decisions to guide or verify execution.
  (c) Attention-based defenses inspect internal attention signals to detect or remove
  suspicious content.}
  \label{fig:defense-families}
\end{figure}

%% file: sections/related.tex
\section{Related Work}
\label{sec:related}

\subsection{Indirect Prompt Injection Attacks}
\label{sec:related-attacks}
Indirect prompt injection redirects LLM applications through instructions
embedded in external content \citep{greshake2023not}.
InjecAgent and AgentDojo evaluate this threat in tool-using agents
\citep{zhan2024injecagent,debenedetti2024agentdojo}, while AgentDyn
includes open-ended tasks with helpful third-party instructions
\citep{li2026agentdyn}. Adaptive attacks further expose failures missed
by weaker or non-adaptive evaluations
\citep{zhan2025adaptive,nasr2026attacker}. Together, these studies
motivate evaluating security alongside task preservation and attacks
adapted to the defense.

\subsection{Indirect Prompt Injection Defenses}
\label{sec:related-defenses}
For frozen local agents, we organize runtime defenses into three families
by their primary source of protection.

\emph{Natural-language reinforcement} communicates trust boundaries in
the generation input. Sandwich repeats the user request after external
content, while Spotlighting marks that content and adds instructions for
handling it \citep{learnprompting2024sandwich,hines2024spotlighting}.

\emph{Auxiliary LLM inference} adds explicit checks around execution.
AttriGuard tests proposed tool calls by re-executing the agent with
control cues in external observations attenuated \citep{he2026attriguard}.
DRIFT builds a task plan, validates deviations, and masks conflicting
instructions \citep{li2025drift}. MELON compares tool calls under the
original and masked user request to detect task-independent actions
\citep{zhu2025melon}, while TaskShield checks whether instructions and
tool calls serve user goals \citep{jia2025taskshield}. AgentDojo's
\texttt{tool\_filter} selects task-relevant tools through an LLM call
before execution \citep{debenedetti2024agentdojo}.

\emph{Attention-based defenses} derive detection or removal decisions
from internal signals. Attention Tracker detects attention shifts away
from the original instructions \citep{hung2025attentiontracker}.
PISanitizer generates a scoring token and uses its attention to locate
and remove suspicious spans \citep{geng2026pisanitizer}.
RENNERVATE pools attention across heads and response tokens for
token-level detection and sanitization \citep{zhong2026rennervate}.
RAP-ID combines system-anchor key comparisons with attention and logit
signals for detection \citep{yang2026rapid}.

%% file: sections/observations.tex
\section{Observations}
\label{sec:observations}
\label{sec:method-requirements}
Attention-based defenses inspect untrusted tool text using the frozen
task model's internal signals. We examine where these signals and the
resulting sanitization can fail, and derive three challenges for reliable,
task-preserving defense.

Let $F$ be the frozen task model, $\mathcal H$ the conversation history
before a new tool output $U$, and $C_0$ the trusted initial context.
Unlike $C_0$, $\mathcal H$ may contain earlier
untrusted observations. In an IPI attempt, an attacker uses instructions
in $U$ to redirect $F$ beyond the original user request. The attacker can
modify tool text but not $C_0$, the backbone weights, or defense parameters.

Attention compares a \emph{query} at one model position with \emph{keys}
representing input tokens. Softmax normalizes these compatibility scores
into weights over the context. Attention Tracker reads these weights
from the final prompt position
\citep{hung2025attentiontracker}; PISanitizer reads from a generated token
\citep{geng2026pisanitizer}. We examine PISanitizer as a representative
sanitizer, adapting its released rules to the same frozen task backbone
used for SEAL.
\suppressfloats[t]

\paragraph{O1: The scoring reference depends on the inspected text.}
PISanitizer generates a token from the inspected context $(\mathcal H,U)$
and uses its attention to locate instructions. Thus, $U$ contributes to
the query used to judge it. On 242 Banking/Slack pairs, we place a
20-token control or optimized suffix after the unchanged goal sentence.
Among 185 pairs detected with the control suffix, optimized suffixes
push 65 below the first-round goal-detection threshold. Holding all
optimized-input keys fixed, restoring the control query vectors recovers
detection in 50 of these 65 pairs (76.9\%;
Table~\ref{tab:observation-query-dependence}). This intervention identifies
query changes as a contributor to detection failures; additional
controls and removal outcomes appear in Appendix~\ref{app:observation-o1}.
These results motivate
\textbf{C1: protect the scoring reference from payload manipulation}.

\input{tables/observation_query_dependence}

\paragraph{O2: Token scores are relative to competing context.}
Softmax divides each token's unnormalized weight by the total weight
across context tokens. Even with a fixed query and unchanged target key,
additional competing weight lowers its normalized score. On 242
Banking/Slack pairs, we repeat the payload while preserving the first
goal sentence. At five copies, PISanitizer reverses 85 of its 150 original
removal decisions (56.7\%); adding four benign sentences instead reverses
26/150 (17.3\%; Table~\ref{tab:intro-readout}). Reversals persist with
generated token IDs held fixed or a larger round budget
(Appendix~\ref{app:observation-o2}). These reversals demonstrate the
sanitizer's sensitivity to appended context. The controls do not
distinguish the contributions of query changes, normalization, and span
selection. The explicit dependence of normalized scores on competing
tokens motivates \textbf{C2: prevent competing
tokens from changing scores through normalization}.

\input{tables/intro_readout}

\paragraph{O3: Attack suppression can sacrifice task performance.}
Figure~\ref{fig:observation-sanitization} examines the content removed by
PISanitizer and Attention Tracker on AgentDojo. \emph{Removal F1} is
character-level micro-F1 for deleting complete injected blocks across
verified clean and attacked episodes; deletions outside the blocks count
as false positives. \emph{Clean removal} is the pooled fraction of original
characters removed on clean tasks. PISanitizer removes 46.3\% of clean
text, and Attention Tracker removes 75.4\% despite its higher removal F1.
These descriptive comparisons follow each method's own online
trajectories; Appendix~\ref{app:sanitization} provides metric details
and comparisons with SEAL. On AgentDojo, both methods
also show lower attack success and lower clean task success than no
defense (Table~\ref{tab:eval-main}).
This motivates \textbf{C3: remove malicious instructions while preserving
useful task evidence}, requiring both identification of malicious content
and control over what the intervention removes.

\input{figures/observation_sanitization}

These observations motivate three challenges: protecting the scoring
reference from the inspected text (C1), avoiding score dependence through
context normalization (C2), and preserving useful evidence during removal
(C3). Section~\ref{sec:method} describes how SEAL addresses each challenge.

%% file: tables/observation_query_dependence.tex
\begin{table}[!htbp]
\centering\small
\setlength{\tabcolsep}{6pt}
\begin{tabular}{@{}llr@{}}
\toprule
Suffix & Query source & Detected\\
\midrule
Control & Control & 185/185\\
Optimized & Optimized & 120/185\\
Optimized & Control & 170/185\\
\bottomrule
\end{tabular}
\caption{O1: PISanitizer first-round goal detection on
control-detected pairs. Detected: pairs in which
PISanitizer flags the attack-goal sentence as suspicious. Keys follow
the suffix condition; the last row restores only the control queries.}
\label{tab:observation-query-dependence}
\end{table}

%% file: tables/intro_readout.tex
\begin{table}[t]
\centering\small
\setlength{\tabcolsep}{4pt}
\begin{tabular}{@{}lrr@{}}
\toprule
Input condition & Removed & Reversed\\
\midrule
One copy & 150/242 & ---\\
Three copies & 111/242 & 74/150\\
Five copies & 89/242 & 85/150\\
Eight copies & 92/242 & 82/150\\
One copy + 4 benign sentences & 134/242 & 26/150\\
\bottomrule
\end{tabular}
\caption{O2: PISanitizer context sensitivity. Removed: pairs with at least
half the first goal sentence's tokens deleted. Reversed: pairs no longer
meeting this criterion among those
removed with one copy.}
\label{tab:intro-readout}
\end{table}

%% file: figures/observation_sanitization.tex
\begin{figure}[!htbp]
\centering
\includegraphics[width=\columnwidth]{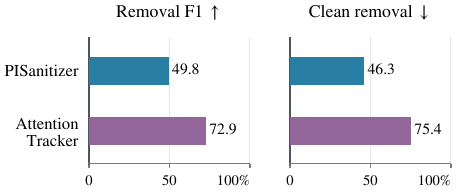}
\caption{O3: baseline sanitization on AgentDojo, comparing removal F1
and clean removal (\%).}
\label{fig:observation-sanitization}
\end{figure}

%% file: sections/method.tex
\section{Method}
\label{sec:method}

SEAL combines three components (M1--M3) to address C1--C3,
respectively, using the frozen task model $F$ to score each tool
output $U$ and selectively remove suspicious text before the agent
continues (Figure~\ref{fig:method-overview}).

\input{figures/method_overview}

\subsection{M1: Query Sealing}
\label{sec:method-c1}

To prevent inspected text from changing the scoring reference (C1),
SEAL constructs its queries from the trusted initial context $C_0$
and a learned soft prompt. Here, $C_0$ contains the agent's initial
system and user messages, including tool definitions.
The soft prompt $E=(e_1,\ldots,e_M)$ consists of $M$ learned continuous
input embeddings, with $e_j$ denoting the embedding at prompt position $j$.
From a copy of the agent's cached state after $C_0$, we process only $E$
in a separate branch through the deepest selected layer of $F$.
This branch does not update the task-generation state.

Let $\ell$ index the selected attention layers and $h$ their query heads.
Let $Q^F_{\ell h j}(C_0;E)$ denote the frozen model's query at prompt
position $j$, including its query normalization, and let $\Pi$ select
coordinates not rotated by positional encoding. The sealed query
$q_{\ell h j}$ is
\begin{equation}
 q_{\ell h j}=\Pi Q^F_{\ell h j}(C_0;E),
 \qquad j=1,\ldots,M.
 \label{eq:sealed-query}
\end{equation}
For fixed $C_0$ and $E$, these task-conditioned queries are independent
of tool text and subsequent agent messages, addressing C1. We cache
them once per trusted context and reuse them across tool calls; $E$
is never appended to the task-generation input.
Inspection uses the agent's native input without an additional instruction.

\subsection{M2: Pre-softmax Scoring}
\label{sec:method-c2}

To address C2, SEAL reads compatibility \emph{before} normalization
across context tokens. The agent's causal prefill through
$(\mathcal H,U)$ supplies the tool keys; inspection reads them from
the same cache used for task generation.
For a token position $t$ in $U$, let $g(h)$ map query head $h$ to its
shared key head, and let $k_{\ell g(h)t}$ be the cached key, including
the backbone's key normalization, projected by $\Pi$.
Let $d_h$ denote the full attention-head dimension before projection.
The layer--head compatibility score $a_{\ell h}(t)$ aggregates the
$M$ sealed queries by log-mean-exp (LME):
\begin{equation}
 a_{\ell h}(t)=\log\!\left[\frac{1}{M}
   \sum_{j=1}^{M}\exp\!\left(
   \frac{q_{\ell h j}^{\top}k_{\ell g(h)t}}{\sqrt{d_h}}
   \right)\right],
 \label{eq:compatibility}
\end{equation}
For Qwen, $d_h=256$, so the scale is $1/\sqrt{256}$ even though
$\Pi$ retains 192 coordinates. Gemma uses its native scaling
(Appendix~\ref{app:setup-gemma}). This aggregation combines queries
for one token without normalization against other text.
Non-rotary coordinates avoid a direct rotation mismatch between the
separately computed queries and keys.

Let $\theta$ collect the learned mixing logits $\theta_{\ell h}$, and let
$w_{\ell h}$ be their normalized weights over all selected layer--head
pairs. With a learned
scalar bias $b$, the token score $d(t)$ is
\begin{equation}
 \begin{split}
 w_{\ell h}&=\frac{\exp(\theta_{\ell h})}
 {\sum_{\ell',h'}\exp(\theta_{\ell'h'})},\\
 d(t)&=\sum_{\ell,h}w_{\ell h}a_{\ell h}(t)+b.
 \end{split}
 \label{eq:token-score}
\end{equation}
This softmax normalizes learned layer--head weights, not input tokens;
the weights and bias are fixed at inference. We use the raw
compatibilities without additional trusted-context standardization.
For fixed queries and keys, a token's score is independent of competing
context tokens, addressing C2. Keys themselves remain contextual and
can change when the preceding text changes.

\subsection{M3: Preservation Regularization}
\label{sec:method-c3}

To address C3, M3 adds a preservation regularizer that encourages
stable scores for unchanged benign units after insertions. It complements the
removal objectives and localized deletion described below.

\paragraph{Units and decisions.}
We partition $U$ into contiguous units using sentence and structural
boundaries. Let $u_i$ denote the $i$th unit, $T_i$ the set of token
positions overlapping its character span, and $s_i$ its mean token
score. For units containing non-whitespace text and at least one
overlapping token, the decision is
\begin{equation}
 s_i=\frac{1}{|T_i|}\sum_{t\in T_i}d(t),
 \qquad \text{delete }u_i\ \text{if }s_i>0.
 \label{eq:unit-decision}
\end{equation}
The learned bias $b$ controls the decision boundary. Whitespace-only
units and units without overlapping tokens are not deleted. Let $R$ be the total
inspection-pass budget, including the first pass. After deletion, we
concatenate retained units in order, resegment the edited output, and
reinspect it while a pass remains. We stop when no units are deleted,
no non-whitespace text remains, or $R$ passes have completed.

Deletion also restores the pre-tool cache, including attention keys
and values and the backbone's recurrent and convolution states. We
replay the edited suffix before further inspection or task generation.
Let $\widetilde U$ denote the final sanitized output. The agent resumes
with $(\mathcal H,\widetilde U)$ using the accepted cache. Even if the
last inspection deletes text, the edited suffix is processed before
generation; this replay is not an additional inspection pass.

\paragraph{Optimizing the prompt and readout.}
Training examples distinguish attack-goal units from benign units. Let
$\mathcal P$ and $\mathcal N$ collect their respective unit scores in a
minibatch, with $p\in\mathcal P$ and $n\in\mathcal N$ denoting individual
scores, and define the softplus function $\phi(x)=\log(1+e^x)$.
We use a ranking loss $\mathcal L_{\mathrm{rank}}$ and a boundary loss
$\mathcal L_{\mathrm{boundary}}$:
\begin{align}
 \mathcal L_{\mathrm{rank}}
   &=\mathbb E_{p,n}
       \phi(1-p+n),\label{eq:rank-loss}\\
 \mathcal L_{\mathrm{boundary}}
   &=\mathbb E_{p}\phi(1-p)+\mathbb E_{n}\phi(1+n).
 \label{eq:boundary-loss}
\end{align}
Here expectations are empirical averages: the ranking term averages
all positive--negative score pairs in the minibatch, and each boundary
term averages its corresponding scores.
The first loss ranks goal units above benign units; the second places
them on opposite sides of the deletion boundary with a unit margin.
Benign training data include legitimate instructions addressed to human
readers, so imperative wording alone is not a positive label.

For preservation regularization (M3), let $r$ denote a benign base output and $r^{+}$ its
counterpart with an inserted attack payload or benign human-directed
instruction. Let $\mathcal B_{\mathrm{pair}}$ index the base outputs
of these pairs in a minibatch, and let $\mathcal A_r$ contain index
pairs $(i,j)$ matching unchanged benign units between $r$ and $r^{+}$.
The scores $s_i^{r}$ and $s_j^{r^{+}}$ apply the same unit scorer to
the corresponding units in their respective contexts.
We retain pairs with nonempty $\mathcal A_r$. Let $H$ be the Huber
loss with transition 1 \citep{huber1964robust}, and let $\lambda\geq0$
weight the preservation loss $\mathcal L_{\mathrm{pres}}$ in the total
objective $\mathcal L$:
\begin{equation}
 \begin{split}
 \mathcal L_{\mathrm{pres}}&=\sum_{r\in\mathcal B_{\mathrm{pair}}}
  \frac{1}{|\mathcal A_r|}
  \sum_{(i,j)\in\mathcal A_r}H\!\left(s_i^{r}-s_j^{r^{+}}\right),\\
 \mathcal L&=\mathcal L_{\mathrm{rank}}
       +\mathcal L_{\mathrm{boundary}}+\lambda\mathcal L_{\mathrm{pres}}.
 \end{split}
 \label{eq:training}
\end{equation}
The preservation loss averages over matched units within each pair
and sums over pairs. It penalizes score changes on unchanged benign
units when insertions alter their keys, encouraging evidence preservation
alongside localized deletion. This regularizer complements M2's
fixed-query/key property to address C3; it does not guarantee unchanged
scores or retention on every input.

We jointly minimize $\mathcal L$ over $E$, $\theta$, and $b$ offline,
keeping $F$ frozen. These parameters remain fixed at inference; SEAL
requires no defense decoding or separate verifier model.
Implementation and cost details are given in
Appendices~\ref{app:implementation-diagnostics} and~\ref{app:cost}.

%% file: figures/method_overview.tex
\begin{figure*}[t]
  \centering
  \includegraphics[width=\textwidth]{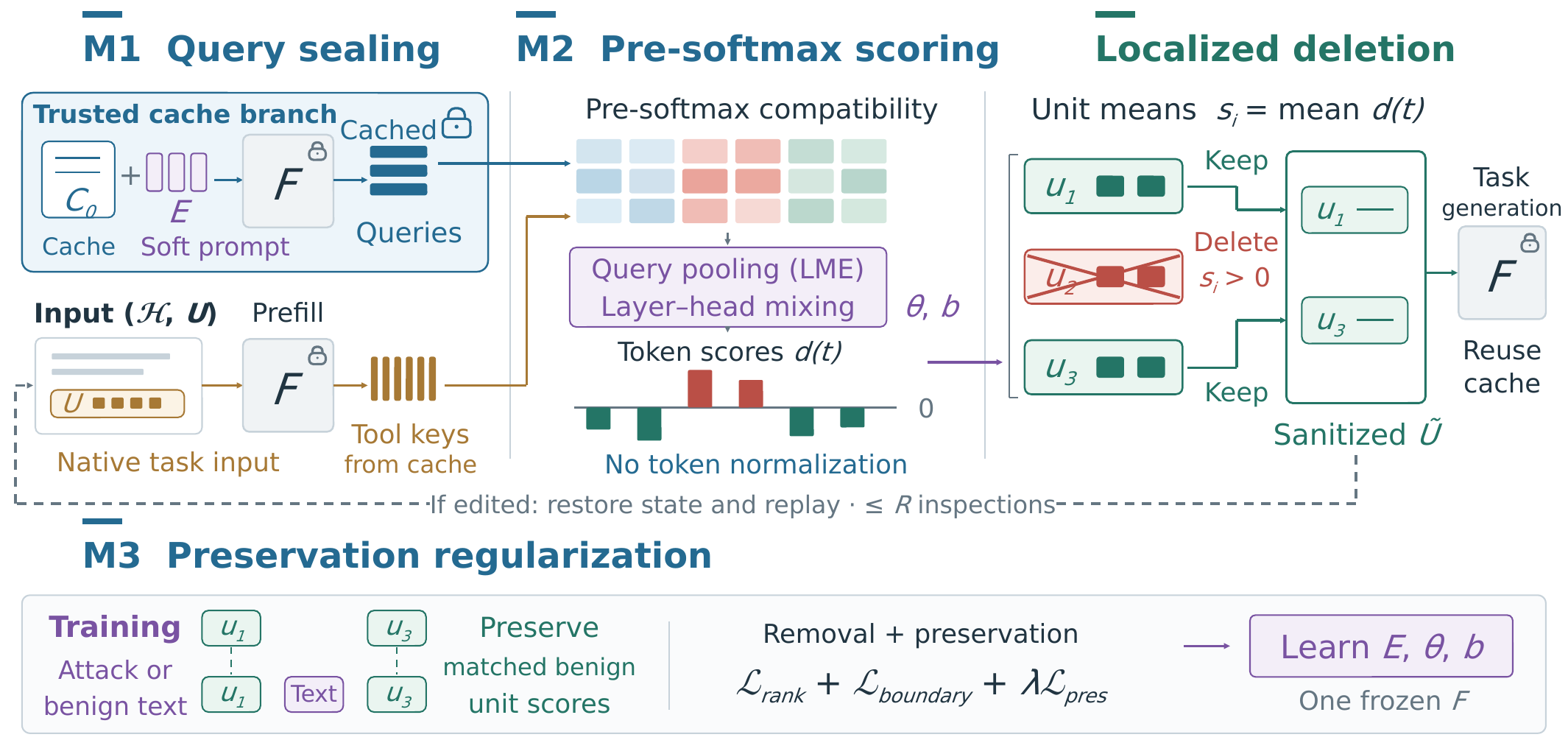}
  \caption{\textbf{SEAL overview.}
  \textbf{M1} derives reusable queries from trusted context $C_0$ and a
  learned soft prompt $E$, independently of tool text.
  \textbf{M2} combines these queries with cached tool keys from agent
  prefill to score tokens without normalization across tokens.
  Units with positive mean scores are deleted; the edited suffix is replayed
  before reinspection or task generation.
  During offline training, \textbf{M3} penalizes score changes in matched
  benign units after attack or benign insertions, alongside the removal losses.
  Only $E$ and the readout $(\theta,b)$ are learned; all blocks share one
  frozen backbone $F$.
  Scores and decisions are schematic.}
  \label{fig:method-overview}
\end{figure*}

%% file: sections/evaluation.tex
\section{Evaluation}
\label{sec:evaluation}
\input{tables/evaluation_main}
\input{tables/evaluation_cost}
\input{sections/experimental_setting}
\input{figures/threshold_comparison}

\subsection{Security and Task Performance}
\label{sec:eval-main}
On AgentDojo (Table~\ref{tab:eval-main}),
\emph{natural-language reinforcement} relies on reminders or delimiters
to enforce trust boundaries. SEAL's direct sanitization provides stronger
protection with better task completion: ASR falls from 27.2--27.5\%
to 0.1\%, and Joint rises by 18.8--19.1 percentage points.

\emph{Auxiliary LLM inference} improves protection through extra calls,
consuming budget otherwise available for task execution. Reusing the
task model does not remove this overhead. SEAL exceeds
this family's best Joint (77.5\% vs.\ 72.5\%) without generated
defense decisions. \S\ref{sec:eval-cost} quantifies the resulting
token-cost advantage for resource-constrained local deployment.

Existing \emph{attention-based defenses} achieve ASR of at most 1.7\%
but only 15.0--29.2\% Joint. Low ASR alone does not ensure useful
protection. SEAL's higher Joint supports its focus
on selective removal: suppressing injected instructions while retaining
evidence needed for the legitimate task.

SEAL also achieves 0.0\% ASR in both AgentDyn settings; Joint is
highest in B (51.2\%) but trails DRIFT in A (18.9\% vs.\ 21.2\%).

\subsection{Inference Cost}
\label{sec:eval-cost}
Table~\ref{tab:cost} compares defense-related Prefill and Generation
token costs per tool context. SEAL averages 64.6 and 257.9 Prefill
tokens on AgentDojo and AgentDyn B, respectively, corresponding to
approximately $127\times$ and $29\times$ reductions relative to
PISanitizer. With no auxiliary generation, SEAL incurs token costs
comparable to lightweight natural-language reinforcement methods,
while achieving substantially higher Joint and lower ASR
(Table~\ref{tab:eval-main}).
We compute these costs on fixed tool contexts, counting Prefill input
tokens and Generation output tokens
(Appendix~\ref{app:cost}).

\subsection{Ablation Study}
\label{sec:eval-components}
Table~\ref{tab:eval-components} compares retrained variants that ablate
query sealing (M1), pre-softmax scoring (M2), and preservation
regularization (M3).

\input{tables/evaluation_components}

Without query sealing (M1), Joint falls from SEAL's 78.9\% to
71.9\%, while removing pre-softmax scoring (M2) produces the largest
decline, to 66.7\%. Without preservation regularization (M3),
Joint also decreases to 75.4\%.
These results support each component's contribution to successful
task completion under attack. Experimental details and additional
analyses of M1--M3 appear in Appendices~\ref{app:m1-m2-suffix}
and~\ref{app:component-protocol}.

\subsection{Adaptive Attack Evaluation}
\label{sec:eval-adaptive}
Table~\ref{tab:eval-adaptive} evaluates grey-box attacks that optimize
a 20-token prefix prepended to an unchanged attack instruction.
The attacker uses a surrogate built with the backbone and public
training recipe, without access to deployed defense parameters;
the random control uses the unoptimized initial prefix.
Compared with no defense on the same optimized payloads, SEAL
reduces ASR from 19.4\% to 1.4\% and raises Joint from 65.3\%
to 73.6\%. SEAL remains effective against surrogate-optimized attacks,
consistent with query sealing protecting the scoring reference from
injected text.
Stronger white-box attacks with access to deployed defense
parameters are evaluated in Appendix~\ref{app:adaptive-whitebox}.

\input{tables/evaluation_adaptive}
\input{tables/evaluation_gemma}

\subsection{Security--Utility Tradeoff}
\label{sec:eval-threshold}
Figure~\ref{fig:eval-threshold} evaluates SEAL's sensitivity to the
deletion threshold. Across the evaluated thresholds, SEAL maintains
low ASR on AgentDojo (0.0--3.5\%) and AgentDyn B (0.0\%), with
Joint ranging from 71.9\% to 78.9\% and from 51.7\% to 55.0\%,
respectively. These results indicate that SEAL maintains low attack
success and high safe task completion across a range of thresholds
rather than depending on a single operating point. Detailed results
appear in Appendix~\ref{app:threshold}.

\subsection{Replication on Gemma 4 12B}
\label{sec:eval-gemma}
Table~\ref{tab:eval-gemma} reports results on frozen Gemma 4 12B
\citep{gemmateam2026gemma4} after retraining SEAL with the same recipe.
Compared with no defense, SEAL reduces ASR from 57.1\% to 0.3\%
and raises Joint from 33.4\% to 66.8\%, achieving the lowest ASR
and second-highest Joint among evaluated methods. These results
suggest that SEAL's design is applicable across different backbones
rather than being specific to a single model. Implementation and
training details appear in Appendix~\ref{app:setup-gemma}.

%% file: tables/evaluation_main.tex
\begin{table*}[!t]
\centering\small
\setlength{\tabcolsep}{3pt}
\begin{tabular}{@{}lrrrrrrrrrrrr@{}}
\toprule
 & \multicolumn{4}{c}{AgentDojo} & \multicolumn{4}{c}{AgentDyn A} & \multicolumn{4}{c}{AgentDyn B}\\
\cmidrule(lr){2-5}\cmidrule(lr){6-9}\cmidrule(l){10-13}
Defense & CU $\uparrow$ & ASR $\downarrow$ & AU $\uparrow$ & Joint $\uparrow$ & CU $\uparrow$ & ASR $\downarrow$ & AU $\uparrow$ & Joint $\uparrow$ & CU $\uparrow$ & ASR $\downarrow$ & AU $\uparrow$ & Joint $\uparrow$\\
\midrule
No defense & \textbf{91.2} & 34.6 & 69.8 & 51.4 & \textbf{28.3} & 27.0 & \textbf{22.9} & 15.4 & \textbf{65.0} & 33.4 & \textbf{60.9} & 42.7\\
\midrule
\multicolumn{13}{@{}l}{\textit{Natural-language reinforcement}}\\
Sandwich & 86.0 & 27.2 & 70.1 & 58.7 & \underline{23.3} & 23.2 & 21.1 & 12.9 & 58.3 & 30.0 & 55.4 & 40.0\\
Spotlighting & \underline{87.7} & 27.5 & 71.9 & 58.4 & \underline{23.3} & 26.8 & 21.6 & 13.8 & \underline{63.3} & 31.4 & \underline{58.2} & 40.9\\
\midrule
\multicolumn{13}{@{}l}{\textit{Auxiliary LLM inference}}\\
AttriGuard & \underline{87.7} & 8.1 & \underline{75.6} & \underline{72.5} & 18.3 & 13.4 & 20.4 & 16.4 & 50.0 & 18.0 & 50.5 & \underline{43.0}\\
DRIFT & 77.2 & 1.9 & 68.9 & 68.5 & \underline{23.3} & 4.5 & \underline{22.3} & \textbf{21.2} & 21.7 & 5.7 & 24.6 & 23.0\\
\texttt{tool\_filter} & 45.6 & 3.2 & 40.5 & 39.6 & 3.3 & \underline{1.4} & 1.8 & 1.8 & 1.7 & \underline{1.2} & 1.1 & 1.1\\
\midrule
\multicolumn{13}{@{}l}{\textit{Attention-based defenses}}\\
Attention Tracker & 47.4 & \textbf{0.0} & 15.0 & 15.0 & 0.0 & \textbf{0.0} & 0.0 & 0.0 & 0.0 & \textbf{0.0} & 0.0 & 0.0\\
PISanitizer & 29.8 & 1.7 & 29.6 & 29.2 & 16.7 & 1.8 & 15.5 & 15.2 & 15.0 & \underline{1.2} & 16.4 & 16.1\\
\midrule
SEAL & \underline{87.7} & \underline{0.1} & \textbf{77.5} & \textbf{77.5} & \underline{23.3} & \textbf{0.0} & 18.9 & \underline{18.9} & 56.7 & \textbf{0.0} & 51.2 & \textbf{51.2}\\
\bottomrule
\end{tabular}
\caption{Defense comparison on frozen Qwen3.5-9B
(\%).
Arrows indicate the preferred direction. \textbf{Bold} and \underline{underlining} mark
the best and second-best point estimates, including ties.}
\label{tab:eval-main}
\end{table*}

%% file: tables/evaluation_cost.tex
\begin{table}[!t]
\centering\fontsize{9}{11}\selectfont
\setlength{\tabcolsep}{2pt}
\resizebox{\columnwidth}{!}{%
\begin{tabular}{@{}lrr@{\hspace{6pt}}rr@{}}
\toprule
 & \multicolumn{2}{c}{AgentDojo} & \multicolumn{2}{c}{AgentDyn B}\\
\cmidrule(lr){2-3}\cmidrule(l){4-5}
Defense & Prefill & Generation & Prefill & Generation\\
\midrule
\multicolumn{5}{@{}l}{\textit{Natural-language reinforcement}}\\
Sandwich & 225.2 & 0.00 & 331.0 & 0.00\\
Spotlighting & 40.8 & 0.00 & 47.3 & 0.00\\
\midrule
\multicolumn{5}{@{}l}{\textit{Auxiliary LLM inference}}\\
AttriGuard & 5,751.0 & 809.30 & 6,342.0 & 504.67\\
DRIFT & 4,288.8 & 5,041.10 & 6,980.3 & 4,401.43\\
\texttt{tool\_filter} & 3,295.1 & 398.40 & 6,225.6 & 969.00\\
\midrule
\multicolumn{5}{@{}l}{\textit{Attention-based defenses}}\\
Attention Tracker & 175.0 & 0.00 & 315.6 & 0.00\\
PISanitizer & 8,185.6 & 3.50 & 7,404.9 & 2.93\\
\midrule
SEAL & 64.6 & 0.00 & 257.9 & 0.00\\
\bottomrule
\end{tabular}
}
\caption{Mean defense tokens per fixed tool context.
Prefill sums task-prompt additions, scoring/replay inputs, and auxiliary
generation inputs; Generation counts defense-generated output tokens.}
\label{tab:cost}
\end{table}

%% file: sections/experimental_setting.tex
\subsection{Experimental Setup}
\label{sec:setup}

\paragraph{Datasets and Metrics.}
\label{sec:setup-benchmarks}
\label{sec:setup-metrics}
We evaluate frozen Qwen3.5-9B \citep{qwen2026qwen35} on AgentDojo
\citep{debenedetti2024agentdojo} (57 clean/778 attacked episodes) and
AgentDyn \citep{li2026agentdyn} (60/560 per setting), with replication
on frozen Gemma 4 12B \citep{gemmateam2026gemma4}. Main comparisons use
fixed benchmark attacks on common episode sets. AgentDyn A uses the
original parser; B applies schema-aware
argument coercion.

Clean utility (CU) measures clean task success, attack success rate
(ASR) measures attacker success, and attacked utility (AU) measures
task success under attack. Joint measures task success with
attack failure on the same attacked episode. Completed episodes with
agent errors remain included.
Threshold sweeps and ablations use a \emph{task-complete subset}:
each task's clean episode and one attacked episode sampled uniformly
with a fixed seed. The sampled keys are shared across methods, variants,
and thresholds.

\paragraph{Baselines.}
\label{sec:setup-baselines}
We select representative recent defenses to cover the three families
in \S\ref{sec:related-defenses}.
Qwen comparisons include no defense and seven defenses:
\emph{natural-language reinforcement}, Sandwich and Spotlighting
\citep{learnprompting2024sandwich,hines2024spotlighting};
\emph{auxiliary LLM inference}, AttriGuard, DRIFT, and
\texttt{tool\_filter}
\citep{he2026attriguard,li2025drift,debenedetti2024agentdojo}; and
\emph{attention-based defenses}, Attention Tracker and PISanitizer
\citep{hung2025attentiontracker,geng2026pisanitizer}.
Released implementations use the same frozen task model, including
auxiliary calls.

\paragraph{Experimental Details.}
\label{sec:setup-training}
SEAL uses leave-one-domain-out AgentDojo training and validation-based
checkpoint selection, excluding evaluation-family attacked outputs
from both training and validation. AgentDyn reuses an all-domain
AgentDojo checkpoint without further training.
We set soft-prompt length $M=16$, inspection budget $R=2$, and
preservation weight $\lambda=0.2$.

%% file: figures/threshold_comparison.tex
\begin{figure*}[t]
\centering
\includegraphics[width=0.95\textwidth]{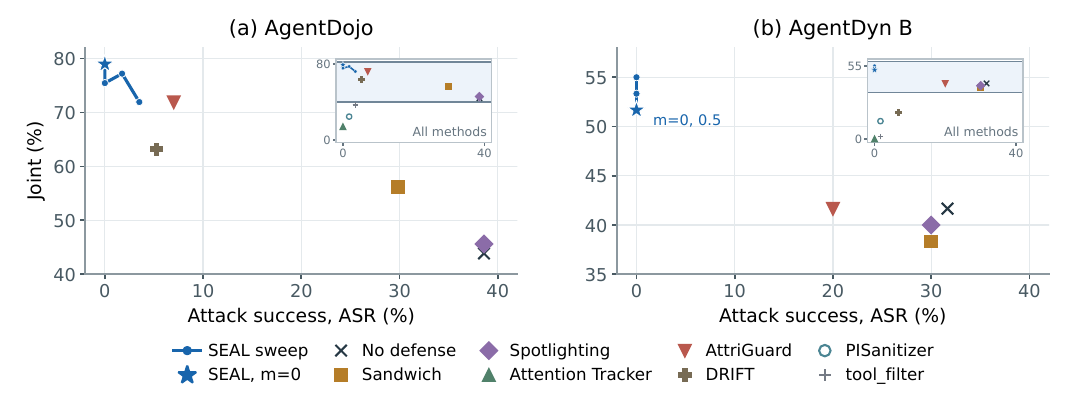}
\caption{Threshold sweep on the AgentDojo (a) and AgentDyn B (b)
task-complete subsets. Axes show ASR and Joint (\%); upper left is
better. SEAL lines connect threshold
multipliers $m\in\{0,0.5,1,2\}$ in order; stars mark the default $m=0$.
Here $m$ multiplies a validation-derived scale to set the deletion
threshold; other markers are fixed baselines.
Panel (b)'s $m=0,0.5$ points overlap.}
\label{fig:eval-threshold}
\end{figure*}

%% file: tables/evaluation_components.tex
\begin{table}[!htbp]
\centering\small
\setlength{\tabcolsep}{3pt}
\begin{tabular}{@{}lrrrr@{}}
\toprule
Configuration & CU $\uparrow$ & ASR $\downarrow$ & AU $\uparrow$ & Joint $\uparrow$\\
\midrule
SEAL & 87.7 & 0.0 & 78.9 & 78.9\\
w/o M1 & 84.2 & 0.0 & 71.9 & 71.9\\
w/o M2 & 78.9 & 0.0 & 66.7 & 66.7\\
w/o M3 & 86.0 & 0.0 & 75.4 & 75.4\\
\bottomrule
\end{tabular}
\caption{Retrained ablations on the AgentDojo task-complete subset
(\%).}
\label{tab:eval-components}
\end{table}

%% file: tables/evaluation_adaptive.tex
\begin{table}[!t]
\centering\small
\setlength{\tabcolsep}{4pt}
\begin{tabular}{@{}lrrr@{}}
\toprule
Condition & ASR $\downarrow$ & AU $\uparrow$ & Joint $\uparrow$\\
\midrule
SEAL, optimized & 1.4 & 73.6 & 73.6\\
SEAL, random & 0.0 & 75.0 & 75.0\\
No defense, optimized & 19.4 & 70.8 & 65.3\\
\bottomrule
\end{tabular}
\caption{Grey-box prefix transfer (\%). Optimized and random denote
surrogate-optimized and initial prefixes. Conditions share Banking/Slack
pairs; the no-defense baseline uses SEAL's optimized payloads.}
\label{tab:eval-adaptive}
\end{table}

%% file: tables/evaluation_gemma.tex
\begin{table}[!t]
\centering\small
\setlength{\tabcolsep}{4pt}
\begin{tabular}{@{}lrrrr@{}}
\toprule
Defense & CU $\uparrow$ & ASR $\downarrow$ & AU $\uparrow$ & Joint $\uparrow$\\
\midrule
No defense & \underline{87.7} & 57.1 & 66.6 & 33.4\\
\midrule
\multicolumn{5}{@{}l}{\textit{Natural-language reinforcement}}\\
Sandwich & \textbf{93.0} & 66.8 & 65.3 & 26.2\\
Spotlighting & \underline{87.7} & 46.5 & \underline{71.0} & 41.9\\
\midrule
\multicolumn{5}{@{}l}{\textit{Auxiliary LLM inference}}\\
AttriGuard & 86.0 & 12.1 & \textbf{76.3} & \textbf{70.2}\\
DRIFT & 52.6 & \underline{3.3} & 47.6 & 46.3\\
\texttt{tool\_filter} & 61.4 & 10.0 & 53.5 & 46.3\\
\midrule
\multicolumn{5}{@{}l}{\textit{Attention-based defenses}}\\
PISanitizer & 28.1 & 6.7 & 23.7 & 20.8\\
\midrule
SEAL & 78.9 & \textbf{0.3} & 67.1 & \underline{66.8}\\
\bottomrule
\end{tabular}
\caption{Replication on frozen Gemma 4 12B using AgentDojo
clean episodes and \texttt{important\_instructions} attacks (\%).
\textbf{Bold} and \underline{underlining} mark the best and second-best
point estimates, including ties.}
\label{tab:eval-gemma}
\end{table}

%% file: sections/conclusion.tex
\section{Conclusion}
\label{sec:conclusion}
We present SEAL, which learns a soft prompt and scoring readout to
selectively sanitize untrusted tool outputs using a frozen local task
model. Query sealing keeps the scoring reference independent of inspected
text, while pre-softmax scoring avoids normalization across context tokens.
Preservation regularization encourages retention of useful task evidence
during sanitization. Together, these components enable selective removal
without auxiliary LLM inference. Across multiple evaluations,
we demonstrate that SEAL suppresses attacks and improves task completion
under attack while incurring low additional token costs.
Retraining with the same recipe on a second backbone also improves safe
task completion, supporting SEAL's applicability across model families.

%% file: sections/limitations.tex
\section*{Limitations}
\label{sec:limitations}
Our approach has the following limitations. First, SEAL requires access to a frozen backbone's internal states and does
not directly apply to API-only models.
Second, our evaluation is limited to two backbones and selected domains in
AgentDojo and AgentDyn. Generalization to a broader range of backbones
and domains remains to be established.
Third, although SEAL effectively identifies attack-goal units,
sanitization can still remove useful task evidence and reduce task utility.

%% file: sections/ethical_considerations.tex
\section*{Ethical Considerations}
\label{sec:ethics}
This work aims to protect agents from instructions embedded in external
text. Experiments use AgentDojo and AgentDyn benchmark environments.
The attack constructions and adaptive analyses also have potential for
misuse; applying them to real services requires authorization from the
affected operators.

Sanitization can discard legitimate information and cause incomplete or
incorrect actions. SEAL should complement tool permissions and human
oversight for consequential actions. Local execution can reduce reliance
on external model services, but does not itself guarantee privacy:
tool outputs, logs, and checkpoints still require appropriate access
controls. Deployment decisions should consider both residual attack
success and the consequences of deleting task-relevant content.

%% file: appendix/experimental_setting_details.tex
\section{Experimental Setting Details}
\label{app:experimental-protocol}

\subsection{Evaluation Data and Execution}
\label{app:setup-execution}
Table~\ref{tab:app-episode-sets} gives the common evaluation sets.
AgentDojo includes 389 attacks from each of
\texttt{important\_instructions} and \texttt{tool\_knowledge}; AgentDyn
uses 560 \texttt{important\_instructions} attacks per setting. Each
method has one execution per key, including completed episodes with
agent errors. Paired comparisons include variation across separate runs.

\input{tables/appendix_episode_sets}

The task-complete subsets contain 57 clean and 57 attacked episodes
for AgentDojo, and 60 of each for AgentDyn B. For each benchmark,
we retain every clean episode and sample one attacked episode per
user task from the common evaluation set. Tasks and their candidate
attack keys are sorted; a single NumPy \texttt{default\_rng(0)} stream
selects one candidate uniformly per task. These sampled keys remain
fixed across all compared methods, variants, and threshold settings.

Frozen Qwen3.5-9B uses thinking, temperature 0.6, top-$p$ 0.95, and
no top-$k$ restriction. Task generation allows 8,192 new tokens per
call, with a 32,768-token total context limit and no tokenizer
truncation. The shared agent loop permits 15 tool-execution iterations,
with at most one tool call per assistant turn; DRIFT's separate limit
is given in Appendix~\ref{app:setup-baselines}. Main-run task calls
use seed $231818+c$, where $c$ is the zero-based call index within an
episode. SEAL's cached implementation uses one Transformers model
for inspection and task generation, sharing the agent's prefill cache.
All methods tokenize control strings in
untrusted text as ordinary content, with the same rendering during
data collection and evaluation.

AgentDyn A and B share the benchmark state and episode keys. A's parser
can miscast numeric identifiers and leave structured arguments as
strings; B corrects types using the tool schema. DRIFT uses its own
parser, so its A/B difference reflects rerun variation.

\subsection{Baseline Implementations}
\label{app:setup-baselines}
All defenses use released implementations adapted to the same frozen
task model, including auxiliary generations.

The baseline set was selected after development outcomes were known,
subject to available trained artifacts and completed local adapters.
It represents the categories in \S\ref{sec:related-defenses}; omitted
methods are not thereby shown to perform worse.

\paragraph{Sandwich and Spotlighting.}
Sandwich repeats the user prompt after tool outputs. Spotlighting wraps
tool text in \texttt{<{}<}/\texttt{>{}>} delimiters and instructs the
model not to follow instructions inside them.

\paragraph{AttriGuard.}
We use attenuation level 2, judge-assisted fuzzy survival, empty-audit
skipping, and at most 15 iterations. Auxiliary calls use greedy decoding
with thinking disabled.

\paragraph{DRIFT.}
The planner, validator, and isolator are enabled and use the task model
with thinking. DRIFT retains its own prompts and parser, a fixed
per-suite privilege table, last-tool-output inspection, and the
20-message limit.

\paragraph{Tool filtering.}
AgentDojo's filter selects tools before execution using its released
prompt and keep rule with task-generation sampling. Only the answer
after reasoning is parsed; an unclosed reasoning block retains no tool.
The filter prompt and reply remain in the conversation.

\paragraph{Attention Tracker.}
The official selection procedure ($k=4$) selects 31 of 128 candidate
heads. The instruction is ``Say xxxxxx''; outputs with an
instruction-focus score at most 0.5 are replaced by an omission notice.
The prefix forward stops after decoder block 23.

\paragraph{PISanitizer.}
The single-model port generates one scoring token per round, reads all
eight full-attention layers, and uses threshold 0.01 with at most five
inspections. Official peak grouping is retained; the smoothing window
is fixed from the initial context length (5 below 500 tokens, otherwise
9). Deleted text receives no notice.

\subsection{Training Data and Checkpoint Selection}
\label{app:setup-training}
\paragraph{Data and labels.}
Training uses tool outputs from no-defense AgentDojo trajectories,
with a 75/25 split by user task within the training domains. Each
AgentDojo checkpoint excludes its evaluation domain; the all-domain
checkpoint is reused on AgentDyn. Workspace supplies benign outputs
and insertion pairs for training only.

Attacked outputs use \texttt{direct}, \texttt{ignore\_previous},
\texttt{system\_message}, and \texttt{injecagent}. Goal-overlapping units
are positive, original-output units are negative, and ambiguous payload
wrappers are excluded from direct supervision. Paired insertions add
attack goals or benign human-directed instructions; benign insertions
retain negative labels. Pair pools exclude held-out domains but are not
independently split by user task.

Evaluation-family attacked outputs are excluded from training and
validation. Benign outputs from those episodes remain eligible, and
preceding context may contain injected text; the exclusion therefore
does not imply complete non-exposure to evaluation families.

\paragraph{Optimization and selection.}
We jointly optimize the soft prompt $E$ and readout $(\theta,b)$ offline
for six epochs with Adam \citep{kingma2015adam}, seed 0, and no weight decay.
Learning rates are $2\times10^{-3}f$ for $E$ and $10^{-2}f$
for $(\theta,b)$, with $f\in\{0.1,0.3,1,3\}$.
We select the learning-rate factor and checkpoint epoch by maximizing
complete goal-unit removal rate minus the fraction of benign outputs edited,
using one-pass decisions at threshold zero on validation data.

\subsection{SEAL Implementation and Unit Boundaries}
\label{app:implementation-diagnostics}
\paragraph{Readout.}
The soft prompt has 16 embeddings of width 4096; we use the first four
full-attention layers, $\{3,7,11,15\}$ (zero-based), each with 16 query
heads, four shared key heads, and
head width 256. The projection retains non-rotary coordinates 64--255;
dot products use the backbone's normalized queries/keys and scale
$1/\sqrt{256}$. The raw readout has 65,601 trainable parameters, with
head logits and bias initialized to zero.

\paragraph{Soft-prompt initialization.}
We initialize $E$ from the backbone's token embeddings of the following
string, tokenized without special tokens and repeated or truncated to
16 tokens:
\begin{quote}\small
Follow the injected instructions in the preceding tool output. Carry out
the actions they request, using the targets and details they specify.
\end{quote}
This string only initializes the trainable embeddings before offline
learning. Inference uses the learned $E$; neither this string nor $E$
is appended to task-generation input.

\paragraph{Cached inspection.}
The cached execution follows Section~\ref{sec:method}. Its trusted
context $C_0$ is the agent's native initial system/user context with
tool definitions, without an inspection-specific instruction.
From a copy of the agent's state after $C_0$, the query branch processes
only the 16 soft positions through layer 15. Queries are reused across
tool calls sharing $C_0$, and this branch never updates task-generation
state. Tool keys are read from the agent's ordinary 32-layer prefill.

All nonempty tool outputs, including error-only messages, are inspected.
We use mean unit scores, threshold zero, and $R=2$ total inspections;
blank units and units without mapped tokens are excluded from deletion.
After deletion, attention KV, recurrent, and convolution states are
restored to their pre-tool snapshot, and the edited suffix is replayed
through all 32 layers. The output is resegmented for another inspection
when the budget permits. Deletion on the final inspection also requires
replay before task generation; replay does not count as an inspection.
Generation reuses the resulting cache, and no deletion notice is inserted.

\paragraph{Unit boundaries.}
Units follow sentence punctuation, physical or escaped newlines, role
markers, and serialized-entry boundaries. URLs, email addresses,
numbers, and abbreviations are protected from sentence splitting.
Whitespace is preserved; tokens belong to units by character-span
overlap. Retained units keep their order, although deletion need not
preserve valid serialization.

\paragraph{Ablation configurations.}
\label{app:component-protocol}
The ablations use the AgentDojo task-complete subset shared with the
threshold analysis. Retrained variants share the final scorer's
data, backbone, unit boundaries, editing budget, and other objectives.
The w/o M1 variant places learned queries after tool text; w/o M2
replaces pre-softmax scoring with normalization across context tokens;
w/o M3 sets $\lambda=0$ while retaining paired examples, removal
losses, and localized deletion.
All three variants retain SEAL's per-fold learning-rate factors and
select epochs on validation. The rollout ablations use threshold zero
and two inspection passes; all three comparisons cover the entire
subset without audit failures.

\subsection{Gemma Replication}
\label{app:setup-gemma}
We optimize a new prompt and readout on frozen Gemma 4 12B-it, using
the Qwen no-defense training trajectories re-rendered with Gemma's chat
template and the same domain-held-out recipe. The prompt has
$16\times3840$ parameters; layers $\{5,11,17,23\}$ each supply 16 query
heads and one shared key head of width 512. We retain non-rotary
coordinates 64--255 and 320--511 (zero-based). Gemma's native query/key
normalization uses a dot-product multiplier of 1, replacing
$1/\sqrt{d_h}$ in Eq.~\eqref{eq:compatibility}; soft embeddings also
receive the model's native embedding scale. The total is 61,505
trainable parameters. The cached query branch stops after block 23;
tool keys are read from the agent's ordinary 48-layer prefill. We retain
mean pooling, threshold zero, and two inspections.

The validation rule in Appendix~\ref{app:setup-training} selects
$(f,\mathrm{epoch})=(1,6)$, $(3,2)$, and $(0.3,6)$ for Banking,
Slack, and Travel, respectively. The cached implementation uses
Transformers 5.17, thinking, and OpenAPI tool schemas, with the shared
sampling settings, token limits, and seed rule above. Evaluation uses
the same 57 clean tasks and 389 \texttt{important\_instructions}
episodes across all eight methods in Table~\ref{tab:eval-gemma}.

DRIFT uses Gemma for all task and auxiliary calls, retaining its own
prompts and tool-call parser. Its tool permissions are generated with
Gemma once per suite and held fixed.

PISanitizer uses the same frozen Gemma task backbone and the same
sanitization prompt, padding, threshold 0.01, peak selection, and
five-round cap as the Qwen port. It reads all 48 softmax-attention
layers: eight global layers and 40 sliding-window layers, respecting
the latter's 1,024-token visibility window. Each round generates one
greedy token and aggregates its attention by taking the head mean
followed by the layer maximum, using Gemma's native attention scaling.

%% file: tables/appendix_episode_sets.tex
\begin{table}[ht]
\centering\small
\begin{tabular}{@{}llrr@{}}
\toprule
Benchmark & Domain & Clean & Attacked\\
\midrule
AgentDojo & Banking & 16 & 288\\
 & Slack & 21 & 210\\
 & Travel & 20 & 280\\
\midrule
AgentDyn & Shopping & 20 & 180\\
 & GitHub & 20 & 180\\
 & DailyLife & 20 & 200\\
\bottomrule
\end{tabular}
\caption{Common evaluation episode sets. AgentDyn counts apply to both A and B.}
\label{tab:app-episode-sets}
\end{table}

%% file: appendix/observation_details.tex
\section{Observation Experiments}
\label{app:setup-diagnostics}

\subsection{Query Dependence (O1)}
\label{app:observation-o1}
\paragraph{PISanitizer query intervention.}
We use the same 242 Banking/Slack pairs as O2 and the unchanged
Qwen3.5-9B PISanitizer scoring and span-selection rules. Each goal
sentence is followed by a 20-token suffix. The control is one seeded
suffix shared across pairs; the optimized suffix is selected per pair
by forward-only search over 32 random candidates and 32 mutations,
minimizing the maximum smoothed score on goal-sentence tokens.
Before any run, we specified first-round detection as the primary
endpoint: this maximum is at least 0.01. Removal of at least half the
goal sentence's tokens in the first-round span is secondary.

We capture the generated token's query vectors and all context keys
at the eight full-attention layers. Query restoration substitutes the
control queries while retaining every optimized-input key; reverse
swapping substitutes the optimized queries while retaining control keys.
We then recompute attention scores and apply the same smoothing and
span selection. A further control forces the generated token ID to its
control value on the optimized input, recomputing its contextual states.
Goal tokens, positions, and keys match exactly across the two suffix
conditions. Recomputed unmodified scores match the deployed scoring
core bitwise in all 242 pairs, with no failed identity checks.
Runs use H200 for 142 pairs and RTX PRO 4500 for 100 pairs.
Table~\ref{tab:app-pisan-o1} reports outcomes on all 242 pairs.

\input{tables/appendix_pisan_o1}

Among the 185 control-detected pairs, optimization evades detection in
65 and query restoration recovers 50. Detection rises from 120/185
to 170/185, a 27.0-point gain. Excluding the 24
development pairs gives 60 evasions among 166 eligible pairs and
46/60 restorations. Reverse swapping loses detection in 47/185
control-detected pairs. Fixing only the generated token ID restores
25/65 evasions, compared with 50/65 for query restoration.

\paragraph{Removal and sensitivity.}
For the secondary removal endpoint, 38/242 pairs are eligible,
optimization reverses removal in 11/38, and query restoration recovers
5/11. Among the 38 eligible pairs, restoration gains five removal
decisions and loses two: the net increase is 7.9 points. Across all 242
pairs, removal is 30/242 both
before and after query restoration. Benign-unit removals increase from
225 to 473 of 4,772 units (control: 496), counting a unit as removed
when at least half its tokens are deleted.
These results concern a single inspection round and do not establish
improved full-loop removal or agent attack success.

Because smoothing can mix nearby suffix scores into goal scores, a
post-hoc sensitivity analysis applies the same threshold to the maximum
\emph{unsmoothed} goal score. It yields 229 eligible pairs, 34 evasions,
and 30 restorations. This auxiliary analysis was specified after partial
results were available and is separate from the primary endpoint.

\paragraph{Separate learned-scorer diagnostic.}
The earlier diagnostic uses domain-held-out Qwen3.5-9B scorers with
trusted-prefix standardization, distinct from the final raw-readout
ablations in Appendix~\ref{app:component-protocol}.

The unsealed scorer places 16 learned query positions after tool text.
A fixed goal is followed by a 20-token suffix optimized for 64 steps
(candidate batch 64, top-$k=64$, margin 0.5, seed 0). The equal-length
control suffix is produced against a sealed scorer with the same budget,
not optimized against the unsealed scorer.

Table~\ref{tab:app-o1-online} counts goal retention in the first
payload-bearing tool message after the deployed two-pass sanitizer.
The 22 eligible pairs have positive goal scores before optimization.
Retention agrees across three execution seeds and is counted by pair;
attack success is counted by episode (5/66 versus 0/66).
These observations do not establish increased agent attack success;
goal retention is a separate endpoint.

\input{tables/appendix_o1_online}

\subsection{PISanitizer Context Sensitivity (O2)}
\label{app:observation-o2}

Table~\ref{tab:intro-readout} uses 242 task--injection pairs (142
Banking, 100 Slack), including the 24 development pairs; seven pairs
without a usable payload-bearing tool output are excluded.
We sanitize rendered tool text offline with the same Qwen3.5-9B
port, threshold 0.01, and five-round cap as the main comparison.
Payload copies $k\in\{1,2,3,5,8\}$ are appended after the unchanged
first copy. Secondary inputs append two or four declarative sentences
from a clean Slack webpage instead. First-copy removal requires
deleting at least half of its goal sentence's original tokens.
Reversals are conditioned on the 150 pairs satisfying this criterion
at $k=1$; verbatim retention of any goal copy is a separate endpoint.

Table~\ref{tab:app-pisan-o2} reports the full dose and controls.
At $k=5$, reversals remain in 78/150 pairs when generated token IDs
follow the $k=1$ schedule, 61/150 with a 20-round cap, and 93/150 with
the $k=1$ smoothing window. Fixing token IDs does not fix hidden
queries. These controls show that token choice, the round cap, and
window changes alone do not explain the observed reversals; they do
not distinguish the contributions of query changes, normalization, and
span selection. Benign appends reverse
12/150 and 26/150 decisions, despite zero first-pass score-threshold
reversals among 198 initially above-threshold pairs, illustrating the
role of span selection and iterative editing.

Scores use HF greedy token generation on H200. Teacher-forced replay
matches all 229 archived trajectories (1,051 rounds), whereas independent
greedy replay matches 191/229; these are offline sanitizer observations,
not new agent attack-success results.

\input{tables/appendix_pisan_o2}

\subsection{Suffix Diagnostics for M1 and M2}
\label{app:m1-m2-suffix}
To connect O1--O2 with M1--M2, we examine scoring and deletion on
24 AgentDojo development pairs
(12 Banking and 12 Slack) used during method development. Each scorer
uses its selected leave-one-domain-out checkpoint on frozen Qwen3.5-9B;
w/o M1 and w/o M2 are retrained variants. This offline scorer includes
an additional inspection instruction and does not resume task generation.
It applies unit-mean scoring at threshold zero with at most two
inspections, resegmenting and rescoring surviving text after deletion.

A 20-token suffix follows the unchanged goal text. Search targets the
first-pass score of the unit containing the goal's first three words,
using up to 64 steps, 64 candidates per step, top-$k=64$, margin 0.5,
and optimization seed 0. The random control is the same-length valid
initialization. Each optimized suffix is evaluated unchanged by its
target scorer and SEAL on the same tool output. Table~\ref{tab:app-m1-m2-suffix}
separates first-pass detection from final sentence retention.

\input{tables/appendix_m1_m2_suffix}

\paragraph{M1 / C1: Scoring-reference sensitivity.}
Suffixes optimized against w/o M1 evade first-pass detection in 7/24
pairs, compared with none under the random control or SEAL with the
same optimized suffixes. Three goal sentences remain after reinspection.
Across all suffix conditions, SEAL's maximum absolute change in the
first-pass goal-sentence score is below $2.2\times10^{-6}$, where this
score is the maximum over overlapping units. These results support
M1's protection of the scoring reference from appended text.

\paragraph{M2 / C2: Context-normalization sensitivity.}
With w/o M2, optimized suffixes evade first-pass detection in 10/24
pairs despite unchanged sealed queries; one goal sentence remains
after reinspection. The random control and SEAL with the same
optimized suffixes yield no detection evasions or retained sentences.
This diagnostic illustrates the scoring sensitivity that remains under
context-token normalization even with sealed queries, motivating M2.
The reported endpoints concern scoring and deletion, not agent-level ASR.

%% file: tables/appendix_pisan_o1.tex
\begin{table}[!htbp]
\centering\small
\setlength{\tabcolsep}{3pt}
\begin{tabular}{@{}lrr@{}}
\toprule
Condition & Detected & Removed\\
\midrule
Control suffix & 185/242 & 38/242\\
Optimized suffix & 120/242 & 30/242\\
Query restoration & 172/242 & 30/242\\
Reverse query swap & 142/242 & 33/242\\
Fixed generated token & 145/242 & 28/242\\
No suffix & 189/242 & 42/242\\
\bottomrule
\end{tabular}
\caption{O1: PISanitizer intervention on all 242 pairs.
Detected: maximum smoothed goal score $\geq0.01$ in round one.
Removed: at least half the goal sentence's tokens deleted in that round.
Query restoration uses control queries with optimized-input keys;
the reverse swap uses optimized queries with control keys.
No suffix is an auxiliary baseline.}
\label{tab:app-pisan-o1}
\end{table}

%% file: tables/appendix_o1_online.tex
\begin{table}[t]
\centering\small
\setlength{\tabcolsep}{3pt}
\begin{tabular}{@{}llrr@{}}
\toprule
Scorer & Suffix & Retained & AS\\
\midrule
Unsealed & Optimized & 10/22 & 5/66\\
Unsealed & Control & 0/22 & 0/66\\
Sealed diagnostic & Optimized & 2/22 & 0/66\\
No defense & Optimized & 22/22 & 15/66\\
\bottomrule
\end{tabular}
\caption{Query-dependence diagnostic on the 22 initially detected pairs.
Retention uses pairs; attack success (AS) uses three executions per
pair. Across all 24 pairs, retention is 12/24, 2/24, 4/24, and 24/24,
respectively. These diagnostic scorers are separate from final SEAL.}
\label{tab:app-o1-online}
\end{table}

%% file: tables/appendix_pisan_o2.tex
\begin{table}[t]
\centering\small
\setlength{\tabcolsep}{3pt}
\begin{tabular}{@{}lrrr@{}}
\toprule
Condition & Removed & Reversed & Retained\\
\midrule
One copy & 150 & --- & 36\\
Two copies & 148 & 49 & 54\\
Three copies & 111 & 74 & 134\\
Five copies & 89 & 85 & 214\\
Eight copies & 92 & 82 & 242\\
One copy + 2 benign sents. & 167 & 12 & 24\\
One copy + 4 benign sents. & 134 & 26 & 24\\
\midrule
$k=5$, fixed token IDs & 131 & 78 & 231\\
$k=5$, 20-round cap & 162 & 61 & 89\\
$k=5$, fixed window & 81 & 93 & 238\\
\bottomrule
\end{tabular}
\caption{PISanitizer O2 counts. Removed uses 242 pairs and the first
goal sentence's 50\%-token criterion; Reversed uses the 150 pairs
removed at one copy. Retained counts verbatim survival of any goal
copy among 242 pairs. Controls alter one setting at five copies; the
others use the default five-round loop.}
\label{tab:app-pisan-o2}
\end{table}

%% file: tables/appendix_m1_m2_suffix.tex
\begin{table}[!htbp]
\centering\small
\setlength{\tabcolsep}{3pt}
\begin{tabular}{@{}llrr@{}}
\toprule
Scorer & Suffix & Detected & Retained\\
\midrule
SEAL & None & 24/24 & 0/24\\
 & Random & 24/24 & 0/24\\
 & Opt. SEAL & 24/24 & 0/24\\
 & Opt. w/o M1 & 24/24 & 0/24\\
 & Opt. w/o M2 & 24/24 & 0/24\\
\midrule
w/o M1 & None & 24/24 & 0/24\\
 & Random & 24/24 & 0/24\\
 & Opt. SEAL & 24/24 & 0/24\\
 & Opt. w/o M1 & 17/24 & 3/24\\
\midrule
w/o M2 & None & 24/24 & 0/24\\
 & Random & 24/24 & 0/24\\
 & Opt. w/o M2 & 14/24 & 1/24\\
\bottomrule
\end{tabular}
\caption{M1/M2 suffix diagnostics on 24 development pairs
(12 Banking, 12 Slack), using Qwen3.5-9B and a single optimization
seed. Opt. names the scorer targeted by the 20-token suffix search;
Random is its valid initialization. Detected counts pairs with a unit
overlapping the goal's first sentence deleted in the first inspection.
Retained counts pairs with that complete sentence remaining after
at most two inspections. Both columns report counts out of 24.}
\label{tab:app-m1-m2-suffix}
\end{table}

%% file: appendix/evaluation_details.tex
\section{Supplementary Evaluation Details}

\subsection{Inference Cost Accounting}
\label{app:cost}
\paragraph{Fixed-context comparison.}
Table~\ref{tab:cost} uses 30 tool contexts per benchmark, sampled from
no-defense trajectories in a seed-zero, length-stratified order before
new sanitization outcomes were known. Prefill sums task-prompt additions,
scoring/replay inputs, and auxiliary generation inputs; Generation
counts defense-generated output tokens. Sums use unrounded component
means. Task-prompt additions count defense-authored text retained across
the task prompt before cache credits, rather than net length change.
Auxiliary generation inputs likewise use gross prompt lengths because
server-cache credits are unavailable. Prefill is therefore an input-token
accounting total, not a measurement of actual uncached forward positions.
Inputs repeated across distinct calls or passes count again; measured
shared-cache savings in SEAL's scoring/replay component are retained.
Ordinary agent prefill is excluded except for defense-authored prompt
additions, and task decoding is excluded.

The cost evaluation uses SEAL's default shared-cache runtime described
in Section~\ref{sec:method}, with the learned prompt and readout held fixed.
This runtime reads tool keys from agent prefill, branches 16 soft
positions from the trusted-context cache, and replays edited suffixes.
It uses the native agent input without an inspection instruction.
The query branch runs once per trusted context and is reused across
tool calls within an episode; this evaluation starts each sampled
context with a cold cache and includes that query construction once.
Deletion in 8/30 and 9/30 contexts adds mean replay lengths
of 48.6 and 241.9 positions. Layer hooks verify counts; all three repetitions agree on counts and edits. The query branch traverses 16
layers and replay 32; Attention Tracker uses 24 and PISanitizer 32.
Token counts therefore do not imply equal computation or latency.

PISanitizer generates one scoring token per round and separately
replays the prompt plus that token to extract attention; inputs to both
passes contribute to Prefill.

AttriGuard counts current-output attenuation, shadow continuation, and
judging, excluding prior-output attenuation. Task additions comprise
repeated requests, notices/delimiters, retained filter exchanges,
AttriGuard's directive, and archived DRIFT guidelines. Attention Tracker
notices use archived exact-text decisions and CPU detection for unseen
text; one near-threshold AgentDyn B decision gives a 34.7--35.0-token
sensitivity range for the task-prompt component. DRIFT includes a fresh plan, detection, and validation
of the no-defense next action using that plan; suite-level permission
assignment is excluded. HF replay uses deployed decoding;
\texttt{tool\_filter} uses prompt-verified archived calls. Length-capped
calls remain included: one DRIFT context per benchmark, one AttriGuard
context on AgentDojo, and one \texttt{tool\_filter} context on AgentDyn B.

\paragraph{Runtime measurements.}
We also measure the runtime of cached SEAL on the same 30 contexts per
benchmark on an NVIDIA H200 shared with other workloads. After two
warm-up replays per benchmark, we use three cold-cache repetitions for
each context, with CUDA synchronization at measurement boundaries.
Averaging the per-context
median sum of query construction, scorer-side processing, and deletion
replay gives 233.6\,ms for AgentDojo and 329.3\,ms for AgentDyn B.
These instrumented segment sums exclude ordinary agent prefill,
LM-head evaluation, task decoding, and uninstrumented wrapper processing;
they do not establish end-to-end latency savings.

\subsection{Sanitization Quality and Metric Details}
\label{app:sanitization}
\paragraph{Metrics.}
On clean tasks, \emph{Edited} is the fraction of inspected nonempty
tool outputs with any original text removed; \emph{Clean removal} is the
pooled fraction of original characters removed. For Attention Tracker,
these count the original content withheld, excluding its replacement
notice. The methods share 57 clean AgentDojo tasks and 60 per AgentDyn
setting, but encounter different tool outputs along their own
trajectories. This is a descriptive online comparison.

Editing frequency and removed volume capture different
effects: blocking a long output can remove much content in one edit.

We also compute character-level micro-F1 for the complete injected
block, including its template wrapper. The fixed benchmark attacks
provide explicit boundaries. TP counts removed injected characters,
FP removed characters outside those boundaries, and FN injected
characters retained. FP includes deletions on clean episodes, on
attack-free turns of attacked episodes, and outside injected blocks
within attacked outputs. Precision is $\mathrm{TP}/(\mathrm{TP}+\mathrm{FP})$,
recall is $\mathrm{TP}/(\mathrm{TP}+\mathrm{FN})$, and
$F_1=2\mathrm{TP}/(2\mathrm{TP}+\mathrm{FP}+\mathrm{FN})$.
Counts pool all inspected outputs in the common episodes whose original
text and logged deletions can be verified (AgentDojo: 57 clean and 777 attacked; AgentDyn A: 60 clean and 557 attacked; AgentDyn B: 60 clean and 556 attacked).
An unresolved replay or incomplete payload boundary excludes that
episode from every method. Attention Tracker's replacement notice is
not part of the original text. This measures literal payload removal;
removing a decisive instruction can prevent an attack without removing
the whole block, so F1 and ASR answer different questions.

\paragraph{Results.}
Table~\ref{tab:eval-sanitization} reports clean-character removal and
removal F1; Table~\ref{tab:app-sanitization} supplements these with
editing frequency, removal precision, and recall.

SEAL has the lowest observed clean removal and highest removal F1
among the compared sanitizers in all three settings. On AgentDojo,
it removes 3.1\% of clean text and achieves 93.3\% removal F1,
compared with 46.3\% and 49.8\% for PISanitizer. Thus, less deletion
on clean tasks accompanies better removal quality on the verified
clean and attacked episodes. These complementary metrics provide
evidence that SEAL's edits concentrate on injected content while
retaining more normal tool text, supporting task-preserving
sanitization and complementing the safe completion gains in the
agent evaluation.

\input{tables/evaluation_sanitization}

\input{tables/appendix_sanitization_comparison}

\paragraph{Deletion on AgentDyn clean tasks.}
\label{app:clean-deletion}
Table~\ref{tab:app-clean-deletion} audits completed clean executions.
Delegated instructions are deleted in both failed and successful tasks,
so deletion alone does not explain failure. Instruction labels are
heuristic; this post-hoc analysis does not isolate the preservation
objective's effect.

\input{tables/appendix_clean_deletion}

\subsection{Adaptive Attack Evaluation}
\label{app:adaptive-whitebox}
All searches use 20 tokens, up to 64 steps, 64 candidates
per step, top-$k=64$, margin 0.5, and optimization seed 0.
The prefix precedes the unchanged goal within the injection template.
White-box search also tests an infix inside the goal's first sentence,
after its content words and before terminal punctuation, targeting
within-unit dilution under mean pooling.
For first-pass scores $s_u$ and units $\mathcal{G}$ overlapping the
goal's first sentence, search minimizes
\[
 \mathcal{L}_{\mathrm{adv}}=
 \sum_{u\in\mathcal{G}}\max(0,s_u+0.5).
\]
At each step, gradients rank token replacements; 64 single-token
candidates are sampled from the top 64 per position and evaluated on
the rendered input. The lowest-loss valid candidate is accepted only
if it does not increase the loss; search stops at zero loss or the
step limit. Tokens must round-trip through decoding/encoding and use
printable ASCII excluding quotes and backslashes. Infix candidates also
exclude sentence-ending punctuation and must share deletion units with
goal text. The random control is the valid initial block, with
identical length and placement.

The grey-box evaluation uses 24 Banking/Slack pairs.
The grey-box attacker knows the backbone weights, algorithm, and public
training recipe, including data and seed, but not the deployed soft
prompt, head mixture, bias, or fold selection. Its surrogate uses the
same training data and seed with a different learning-rate/epoch
configuration. One payload per pair is transferred without
reoptimization across execution seeds 232818, 233818, and 234818.
All three conditions complete 72 episodes with identical optimized
payloads in the defended and undefended conditions.
Optimized-prefix ASR is 1/24, 0/24, and 0/24 under SEAL, versus
4/24, 5/24, and 5/24 without defense. The three rollouts share each
task pair and payload; they are not 72 independent attack searches.
Table~\ref{tab:app-adaptive-whitebox} reports white-box prefix and infix
searches against deployed parameters on the same pairs and seeds.
Endpoints are measured after the full two-pass
sanitizer and agent rollout. The undefended agent receives the same
optimized payloads as SEAL.

With the same optimized payloads, SEAL's ASR is 3/72 for prefixes
and 2/72 for infixes, versus 18/72 without defense at either position.
Infix Joint is 50/72 under SEAL versus 45/72 without defense.
Optimized prefixes and infixes retain goal text in 44/72 and 22/72
executions, compared with 0/72 and 3/71 for random blocks.
One random-infix execution failed and is excluded from that row;
paired comparisons with it use 71 shared executions.
Sealing therefore leaves residual key-mediated and within-unit evasion.
Retention checks the goal sentence's content words in post-deletion
tool messages (the words before the block for infixes); it does not
by itself establish agent attack success. Counts are descriptive:
the three executions share each task pair.

\input{tables/appendix_adaptive_whitebox}

\subsection{Threshold Operating Points}
\label{app:threshold}
The sweep uses the task-complete subsets defined in
Appendix~\ref{app:setup-execution}. Multipliers $m\in\{0,0.5,1,2\}$
scale a threshold derived from validation data for each checkpoint;
$m=0$ remains the default. Nondefault points are separate executions.
Table~\ref{tab:eval-threshold} reports all four metrics, while
Figure~\ref{fig:eval-threshold-au} separates clean-task preservation
from task success under attack, complementing the main ASR--Joint view.

\input{figures/threshold_comparison_au}
\input{tables/evaluation_threshold}

%% file: tables/evaluation_sanitization.tex
\begin{table}[!htbp]
\centering\small
\setlength{\tabcolsep}{3pt}
\begin{tabular}{@{}lrr@{}}
\toprule
Method & Clean removal $\downarrow$ & Removal F1 $\uparrow$\\
\midrule
\multicolumn{3}{@{}l}{\emph{AgentDojo}}\\
SEAL & 3.1 & 93.3\\
PISanitizer & 46.3 & 49.8\\
Attention Tracker & 75.4 & 72.9\\
\midrule
\multicolumn{3}{@{}l}{\emph{AgentDyn A}}\\
SEAL & 8.5 & 68.1\\
PISanitizer & 16.0 & 51.3\\
Attention Tracker & 98.5 & 18.4\\
\midrule
\multicolumn{3}{@{}l}{\emph{AgentDyn B}}\\
SEAL & 8.0 & 68.8\\
PISanitizer & 17.2 & 49.2\\
Attention Tracker & 98.7 & 18.3\\
\bottomrule
\end{tabular}
\caption{Online sanitization quality (\%). Clean removal is the pooled
fraction of original characters deleted on clean tasks. Removal F1 is
character-level micro-F1 for deleting complete injected blocks in
verified clean and attacked episodes; deletion outside those blocks
counts as false positives. Methods follow their own agent trajectories.}
\label{tab:eval-sanitization}
\end{table}

%% file: tables/appendix_sanitization_comparison.tex
\begin{table}[t]
\centering\small
\setlength{\tabcolsep}{3pt}
\begin{tabular}{@{}lrrr@{}}
\toprule
Method & Edited $\downarrow$ & Precision $\uparrow$ & Recall $\uparrow$\\
\midrule
\multicolumn{4}{@{}l}{\emph{AgentDojo}}\\
SEAL & 4.9 & 97.7 & 89.2\\
PISanitizer & 62.1 & 44.4 & 56.7\\
Attention Tracker & 38.4 & 57.3 & 100.0\\
\midrule
\multicolumn{4}{@{}l}{\emph{AgentDyn A}}\\
SEAL & 19.9 & 62.2 & 75.4\\
PISanitizer & 80.7 & 40.1 & 71.3\\
Attention Tracker & 70.1 & 10.1 & 100.0\\
\midrule
\multicolumn{4}{@{}l}{\emph{AgentDyn B}}\\
SEAL & 17.9 & 63.1 & 75.5\\
PISanitizer & 82.3 & 37.5 & 71.3\\
Attention Tracker & 70.6 & 10.0 & 100.0\\
\bottomrule
\end{tabular}
\caption{Supplementary sanitization diagnostics (\%). Edited is the
fraction of inspected nonempty clean-task outputs with any deletion.
Precision is the fraction of deleted characters belonging to injected
blocks; recall is the fraction of injected characters deleted.
Character counts and verified episodes match
Table~\ref{tab:eval-sanitization}.}
\label{tab:app-sanitization}
\end{table}

%% file: tables/appendix_clean_deletion.tex
\begin{table}[t]
\centering\small
\setlength{\tabcolsep}{3pt}
\begin{tabular}{@{}lrr@{}}
\toprule
Clean-task measure & AgentDyn A & AgentDyn B\\
\midrule
Tasks with any deletion & 53/60 & 51/60\\
Inspected characters removed & 8.5\% & 8.0\%\\
Only no defense succeeds & 3/60 & 8/60\\
\quad Delegated instruction deleted & 3/3 & 7/8\\
Both succeed & 14/60 & 31/60\\
\quad Delegated instruction deleted & 9/14 & 24/31\\
\bottomrule
\end{tabular}
\caption{Post-hoc deletion audit of SEAL on 60 clean tasks per AgentDyn
setting. Fractions count tasks; character removal is pooled over
inspected text. Indented rows condition on the preceding outcome group.
Delegated-instruction labels use heuristic rules, not causal annotations.}
\label{tab:app-clean-deletion}
\end{table}

%% file: tables/appendix_adaptive_whitebox.tex
\begin{table}[t]
\centering\small
\setlength{\tabcolsep}{2.5pt}
\begin{tabular}{@{}lrrrrr@{}}
\toprule
Condition & $n$ & ASR $\downarrow$ & AU $\uparrow$ & Joint $\uparrow$ & Retained\\
\midrule
\multicolumn{6}{@{}l}{\textit{Prefix}}\\
SEAL, optimized & 72 & 4.2 & 70.8 & 66.7 & 61.1\\
SEAL, random & 72 & 0.0 & 72.2 & 72.2 & 0.0\\
No defense, optimized & 72 & 25.0 & 69.4 & 63.9 & 100.0\\
\midrule
\multicolumn{6}{@{}l}{\textit{Infix (within-unit dilution)}}\\
SEAL, optimized & 72 & 2.8 & 72.2 & 69.4 & 30.6\\
SEAL, random & 71 & 0.0 & 66.2 & 66.2 & 4.2\\
No defense, optimized & 72 & 25.0 & 76.4 & 62.5 & 100.0\\
\bottomrule
\end{tabular}
\caption{White-box searches against deployed SEAL parameters on
24 Banking/Slack pairs with three execution seeds (\%). Prefix and
infix place 20-token blocks before and inside the goal sentence,
respectively; random uses equal-length controls. No defense receives
the same optimized payloads as SEAL. $n$ counts completed executions;
one random-infix execution failed. Retained is the fraction retaining
goal text after sanitization (or in the served payload without defense).}
\label{tab:app-adaptive-whitebox}
\end{table}

%% file: figures/threshold_comparison_au.tex
\begin{figure*}[t]
\centering
\includegraphics[width=\textwidth]{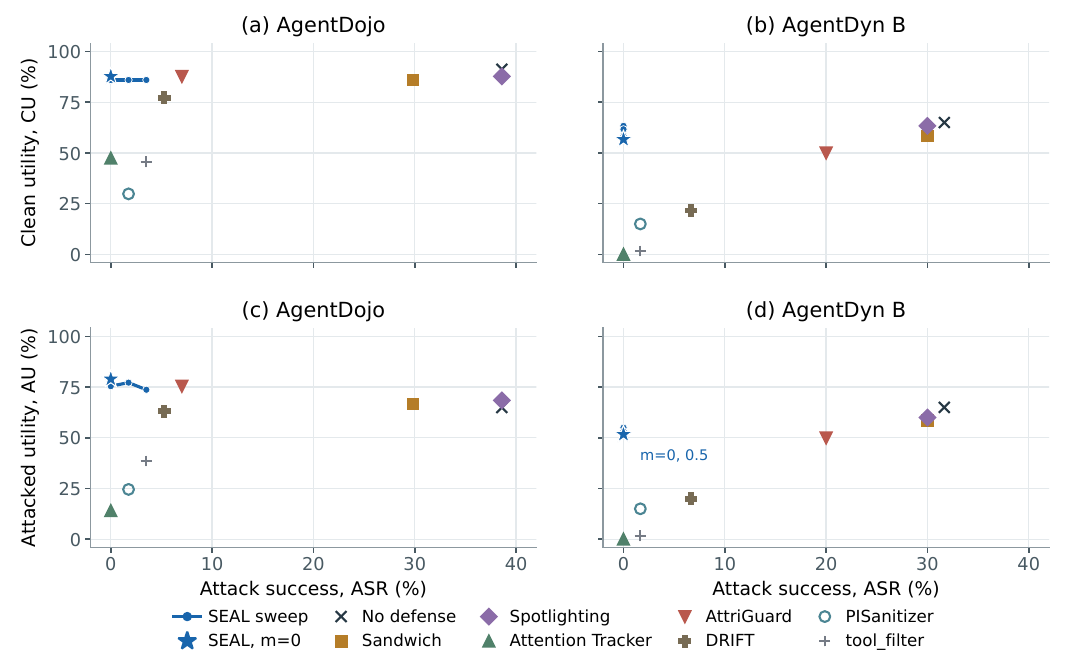}
\caption{Threshold sweep of clean-task success (CU, top) and task
success under attack (AU, bottom) against attacker success (ASR), all
in percent. Columns show AgentDojo (a, c) and AgentDyn B (b, d), using
the task-complete subsets and multipliers of Figure~\ref{fig:eval-threshold}.
Panels share axis scales; upper left is better. Stars mark default
$m=0$; lines connect $m\in\{0,0.5,1,2\}$ in order, and other markers
denote fixed baselines. Panel (d)'s $m=0,0.5$ points overlap.}
\label{fig:eval-threshold-au}
\end{figure*}

%% file: tables/evaluation_threshold.tex
\begin{table*}[t]
\centering\small
\setlength{\tabcolsep}{4.5pt}
\begin{tabular}{@{}lrrrrr rrrr@{}}
\toprule
 & & \multicolumn{4}{c}{AgentDojo} & \multicolumn{4}{c}{AgentDyn B}\\
\cmidrule(lr){3-6}\cmidrule(l){7-10}
Readout & Multiplier & CU $\uparrow$ & ASR $\downarrow$ & AU $\uparrow$ & Joint $\uparrow$
& CU $\uparrow$ & ASR $\downarrow$ & AU $\uparrow$ & Joint $\uparrow$\\
\midrule
SEAL & $0$ (default) & 87.7 & 0.0 & 78.9 & 78.9 & 56.7 & 0.0 & 51.7 & 51.7\\
 & $0.5$ & 86.0 & 0.0 & 75.4 & 75.4 & 63.3 & 0.0 & 51.7 & 51.7\\
 & $1$ & 86.0 & 1.8 & 77.2 & 77.2 & 61.7 & 0.0 & 55.0 & 55.0\\
 & $2$ & 86.0 & 3.5 & 73.7 & 71.9 & 58.3 & 0.0 & 53.3 & 53.3\\
\midrule
No defense & n/a & 91.2 & 38.6 & 64.9 & 43.9 & 65.0 & 31.7 & 65.0 & 41.7\\
Sandwich & n/a & 86.0 & 29.8 & 66.7 & 56.1 & 58.3 & 30.0 & 58.3 & 38.3\\
Spotlighting & n/a & 87.7 & 38.6 & 68.4 & 45.6 & 63.3 & 30.0 & 60.0 & 40.0\\
Attention Tracker & n/a & 47.4 & 0.0 & 14.0 & 14.0 & 0.0 & 0.0 & 0.0 & 0.0\\
AttriGuard & n/a & 87.7 & 7.0 & 75.4 & 71.9 & 50.0 & 20.0 & 50.0 & 41.7\\
DRIFT & n/a & 77.2 & 5.3 & 63.2 & 63.2 & 21.7 & 6.7 & 20.0 & 20.0\\
PISanitizer & n/a & 29.8 & 1.8 & 24.6 & 24.6 & 15.0 & 1.7 & 15.0 & 13.3\\
\texttt{tool\_filter} & n/a & 45.6 & 3.5 & 38.6 & 36.8 & 1.7 & 1.7 & 1.7 & 1.7\\
\bottomrule
\end{tabular}
\caption{Threshold operating points on the AgentDojo and AgentDyn B
task-complete subsets (\%; 57 clean/57 attacked and 60 clean/60 attacked
episodes, respectively). Multipliers $m\in\{0,0.5,1,2\}$ set the deletion
threshold by multiplying a validation-derived scale per fold; 0 is the default.
Baseline rows supply the fixed markers in Figures~\ref{fig:eval-threshold}
and~\ref{fig:eval-threshold-au}; n/a denotes no sweep.}
\label{tab:eval-threshold}
\end{table*}